\documentstyle[preprint,aps]{revtex}
\newcommand{\be}{\begin{equation}} \newcommand{\ee}{\end{equation}} 
\newcommand{\bea}{\begin{eqnarray}}\newcommand{\eea}{\end{eqnarray}}

\begin{document}
\draft
\preprint{MRI-PHY/P/970928, IOP-BBSR/97-41, quant-ph/yymmxxx}
\title{\bf {Supersymmetry, Shape Invariance and Solvability of 
        $A_{N-1}$ and $BC_{N}$ Calogero-Sutherland  Model}}
\author{Pijush K. Ghosh$^{a,1}$, Avinash Khare $^{b,2}$ and
M. Sivakumar$^{c,3}$}
\address{$^a$ The Mehta Research Institute of
Mathematics \& Mathematical Physics,\\
Chhatnag Road, Jhusi,
Allahabad-221 506, INDIA.\\
$^b$ Institute of Physics,
Sachivalaya Marg,
Bhubaneswar-751 005, INDIA.\\
$^c$ School of Physics,
University Of Hyderabad,
Hyderabad-500 046, INDIA.}

\footnotetext {$\mbox{}^1$ Electronic address: 
pijush@mri.ernet.in }  
\footnotetext {$\mbox{}^2$ Electronic address:khare@iop.ren.nic.in}
\footnotetext {$\mbox{}^3$ Electronic address:mssp@uohyd.ernet.in}
\maketitle
\begin{abstract} 
Using the ideas of supersymmetry and shape invariance we re-derive the
spectrum of the $A_{N-1}$ and $BC_N$ Calogero-Sutherland model.  We
briefly discuss as to how to obtain the corresponding eigenfunctions.  We
also discuss the difficulties involved in extending this approach to the
trigonometric models.
\end{abstract}
\narrowtext

\newpage

\section{Introduction}
	
In recent years, supersymmetric quantum mechanics (SUSY QM) has provided a
deeper understanding of the exact solvability of several well known
potentials in 1-dimensional QM. In particular, using the ideas of shape
invariance (SI), it provides a procedure for getting the spectrum, the
eigenfunctions and the S-matrix (i.e.  the reflection and transmission
coefficients) algebraically [1].  There also exist interesting connections
between SUSY QM and soliton solutions. Despite these (and more)
interesting developments in SUSY QM for one particle system in one
dimension, so far, not many of these results could be extended either for
$N$-particle systems in one dimension or for one particle systems in more
than one dimension.

In recent times there is a revival of interest in the  $N$-body problems
in one dimension with inverse square interaction which were introduced and
studied by Calogero [2] and developed by Sutherland [3] and others [4].
These models have several interesting features, like exact solvability,
classical and quantum integrability and also have interesting applications
in several branches of physics [5,6]. Apart from the well known
translational invariant inverse square interaction models, referred to as
$A_{N - 1}$ Calogero-Sutherland Model (CSM), there also exist
generalizations of this model, but without  the translational invariance,
referred to as $BC_{N}$, $B_N$, $D_N$ models. These nomenclatures refer to
the relationship of these models to the root system of the classical Lie
group. It might be added here that these models also share with $A_{N -
1}$ CSM, features like exact solvability, and integrability and have also
found application in certain physical systems.

The purpose of this note is to enquire if the ideas of one dimensional
SUSY QM could be extended to the $N$-particle case. In particular, whether
the spectrum of the celebrated Calogero and other models could be obtained
algebraically by using the ideas of SI and SUSY QM.  The first step in
that direction was taken recently by Efthimiou and Spector [7] who showed
that the well known Calogero model (also termed as $A_{N-1}$ CSM) exhibits
SI. However, they were unable to obtain the spectrum algebraically. This
is because using SUSY they were unable to relate the eigenspectra of the
two SUSY partner potentials.  In this paper we demonstrate that using SUSY
QM, SI and exchange operator formalism [8], the spectrum of the rational
$A_{N-1}$ CSM, and also of all its generalizations like $B_N$, $D_N$ and
$BC_N$ can be obtained algebraically.  It is worth mentioning that the SI
in our case is somewhat different from that of Efthimiou and Spector [7].
So far as we are aware off, this is the first instance when an
$N$-particle quantum system has been solved using the techniques of SUSY
QM and SI.

The plan of the paper is the following. We briefly review the ideas of one
dimensional SUSY QM in Sec. II with the main emphasis on solvability using
SI.. In Sec. II.A, we apply these ideas to the Calogero model, i.e., the
rational $A_{N-1}$ model.  We show that the spectrum of such a model can
be derived using the ideas of SUSY QM, SI and the exchange operator
formalism[8]. We also briefly discuss as to how to obtain the
corresponding eigen-functions. In Sec. II.B, we treat the rational $BC_N$
model, a translationally non-invariant system, in the same spirit.  The
full spectrum is obtained and the method for obtaining the exact
eigen-functions is explicitly spelled out. It is also shown in this
section that the $BC_N$ model posses SI even if the exchange operator
formalism is not employed. This is a generalization of Efthimiou et al's
work [7] on $A_{N-1}$ model to the $BC_{N}$ case.  Finally, in Sec. III,
we summarize our results and discuss the possible directions to be
followed in order to have a viable formalism of many-body SUSY QM.  We
also point out the difficulties involved in extending these results to the
trigonometric case.  In Appendix we show that the $BC_N$ trigonometric
model is also shape invariant.

\section{SUSY, SI and Solvability}

It may be worthwhile to first mention the key steps involved in obtaining
the eigen-spectrum of a one body problem by using the concepts of SUSY QM
and SI.  One usually defines the SUSY partner potentials $H_1$ and $H_2$
by
\be\label{1.1} 
H_1= A^{\dag} A \, , \ H_2 = A A^{\dagger}~~,
\ee
\noindent where ($\hbar = 2m =1$)
\be\label{1.2}
A = {d \over dx} + W(x)~,~ 
A^{\dag} = -{d \over dx} + W(x)~.
\ee
In the case of unbroken SUSY, the ground state wave function is given in 
terms of the superpotential $W(x)$ by, 
\be\label{1.3} 
\psi_0 (x) \propto e^{-\int^{x} W(y)dy}~~,
\ee
\noindent while the  
energy eigenvalues and the wave functions of $H_1$ and $H_2$
are related by, 
$(n=0,1,2,...)$
\be\label{1.4}
E_n^{(2)} = E_{n+1}^{(1)}~~, \hspace{.2in} E_0^{(1)} = 0~~,
\ee
\be \label{1.5}
\psi_n^{(2)} = [E_{n+1}^{(1)}]^{-1/2} A \psi_{n+1}^{(1)}~~, \ \ \ \
\psi_{n+1}^{(1)} = [E_{n}^{(2)}]^{-1/2} A^{\dag}  \psi_{n}^{(2)}~~.
\ee

Let us now explain precisely what one means by SI.
If the pair of SUSY partner Hamiltonians $H_1,H_2$
defined above are similar in shape and differ only in the
parameters that appear in them, then they are said to be SI.
More precisely, if the partner Hamiltonians $H_{1,2}(x;a_1)$ satisfy the
condition,
\be \label{1.6}
H_2(x;a_1) = H_1(x;a_2) + R(a_1),
\ee
where $a_1$ is a set of parameters, $a_2$ is a function of $a_1$ (say
$a_2=f(a_1)$) and the remainder $R(a_1)$ is independent of $x$, then
$H_{1}(x;a_1)$ and $H_{2}(x;a_1)$ are said to be SI. 
The property of SI permits an immediate analytic determination of the
energy eigenvalues, eigenfunctions and the scattering matrix [1]. In
particular the eigenvalues and the eigenfunctions of $H_1$ are given by
($n = 1,2,...$)
\be \label{1.7}
E^{(1)}_n (a_1) = \sum^n_{k=1} R(a_k)~, \hspace {.2in}  E^{(1)}_0 (a_1)=0~~,
\ee
\be\label{1.8}
\psi^{(1)}_n (x;a_1) \propto A^{\dag}(x;a_1)A^{\dag}(x;a_2)....A^{\dag}(x;a_n)
\psi^{(1)}_0 (x;a_{n+1})~,
\ee
\be\label{1.9}
\psi^{(1)}_0 (x;a_1) \propto e^{-\int^{x} W(y;a_1) dy}~~.
\ee
	
\subsection{Rational $A_{N-1}$ Calogero Model}

We now apply the exchange operator formulation to the rational $A_{N-1}$ CSM.
The Hamiltonian of the rational $A_{N-1}$ CSM is given by 
\be\label{9}
H_{CSM} =  {\sum_{i}{\frac{1}{2}} {p_{i}^2}}  + l(l \mp 1) \sum_{i < j} 
{(x_{i} - x_{j})}^{-2} + \omega {\sum_{i} {\frac{1}{2}} {{x_{i}^2}}} \, .
\ee
\noindent 
The sign $\mp$ in (\ref{9})
refers to the fact that $H$ acts on completely anti-symmetric or symmetric
functions, respectively.
Let us now define an operator $D_i$
\be\label{10}
D_{i} = -i \partial_{i} + il \sum_{j}^\prime {(x_i -x_j)}^{-1} M_{ij} ,
\ee
\noindent known as the Dunkl operator in the literature. Hereafter
$\prime$ means $i=j$ is excluded in the summation. The exchange operator
$M_{ij}$
have the following properties [8]
\bea
M_{ij}^2=1~, \ \ \ M_{ij}^\dagger=M_{ij}~,
\ \ \ M_{ij} \psi^{\pm} = \pm \psi^{\pm}~,\nonumber \\
M_{ij} D_{i} = D_{j} M_{ij}~, \ \
M_{ij} D_k = D_k M_{ij}~, \ \ k \neq i,j~~,\nonumber \\
M_{ijk}=M_{ij} M_{jk}~, \ \ M_{ijk}=M_{kij}=M_{jki}~,
\eea
\noindent where $\psi^{\pm}$ is a(an) symmetric(antisymmetric) function.
Note that the Dunkl operator is hermitian by construction
and $[D_i, D_j]=0$. If we now define
\be\label{11}
a_i = D_i - i \omega x_i \, , \ 
{a_i}^\dagger = D_i +i \omega x_i \, ,
\ee
then it is easy to see that
\be\label{12}
{[ a_i, {a_j}^\dagger]} = 2 w {\delta_{ij}}(1 + l{\sum_{k}^{\prime}M_{ik}})
	     - 2 (1-{\delta_{ij}}) l w M_{ij} \, .
\ee
\noindent Let us now consider the SUSY partner potentials 
$H$ and $\tilde {H}$ defined by
\be\label{13}
H = \frac{1}{2}\sum_{i}{a_i}^\dagger a_i~,\ \ \ 
\tilde {H} =\frac{1}{2} \sum_{i}a_i {{a_i}^\dagger} \, . 
\ee
\noindent Using eqs. (\ref{10}) and (\ref{11}) it is easily shown that 
\be\label{14}
H_{CSM} = H + E_0^{CSM}~, \ \ \
E_0^{CSM} = [\frac{N}{2} \mp \frac{l}{2} N (N-1)] \omega~..
\ee
\noindent Thus, by construction, the ground state energy of $H$ is zero.

Using eqs. (\ref{10}) and (\ref{11}) it is easily shown that if  
$\psi$ is the eigenstate of $H$ with eigenvalue $E (>0)$, then 
$A_1 \psi$ is the eigenstate of $\tilde H$ with eigenvalue $E+ \delta_1$ i.e. 
\be\label{16}
\tilde{H} (A_1 \psi)=[E+ \delta_1](A_1 \psi),
\ee
where,
\be\label{17}
A_1 ={\sum_{i}} a_{i}~, \ \ \
\delta_1 = [(N-1) \pm l  N(N-1)] \omega~~.
\ee
\noindent 
Similarly, if $\tilde {\psi}$ is the eigenfunction of $\tilde H$ with 
eigenvalue $\tilde E$, then $A_1^{\dag}\psi$ is the eigenfunction of $H$ 
with eigenvalue $\tilde{E} - \delta_1$ i.e.
\be\label{19}
H({{A}_1^\dagger}\tilde {\psi}) =[\tilde {E}-\delta_1] ({{A}_1^\dagger} 
\tilde{\psi}).
\ee
\noindent This proves one to one correspondence between the non-zero
energy eigen values of $H$ and $\tilde {H}$. 
Thus it follows from here that the energy eigenvalues and eigenfunctions 
of the two partner Hamiltonians $H$ and $\tilde H$ 
are related by 
\be\label{20}
\tilde{E}_n = E_{n+1} +\delta_1 \, ,~~ E_0 = 0 \, , \ \ n = 0,1,2,...
\ee 
\be\label{21}
\tilde{\psi}_n = {A_1 \psi_{n+1} \over \sqrt {E_{n+1} +\delta_1}}~~,
~\psi_{n+1} = {A_1^{\dag} \tilde{\psi}_n \over \sqrt{E_{n+1}}}~~.
\ee
\noindent Note that $\delta_1$ vanishes for $N=1$ and we
recover the usual results of SUSY QM with one degree of freedom.
It is worth noting that unlike the case of one dimensional QM, 
in this case the (positive) energy levels of $H$ and $\tilde{H}$ 
are not degenerate. 

Using eqs.. (\ref{10}) and (\ref{11}) it is also easily shown that $H$ and 
$\tilde{H}$ satisfy the shape invariance condition 
\be\label{23}
\tilde{H} (\{x_i\} ,l)  = H (\{x_i\}, l) +R(l),
\ee
\noindent where
\be\label{24}
R(l) = [ N \pm l N(N-1)] \omega = \omega + \delta_1~~.
\ee
\noindent 
As a result, using the formalism of SUSY QM, and the relation between
$E_{n+1}$ and $\tilde{E_n}$ as given by eq. (\ref{20}), 
the spectrum of H is given by
\be \label{25}
{E_{n}} =\sum_{i}R(l_{i}) - n\delta_1~.
\ee
\noindent Note that in this particular case all $l_i$ are 
identical so that using $\delta_1$ and $R(l)$ as given by eqs. (\ref{17}) 
and (\ref{24}), the spectrum turns out to be
\be\label{26}
E_{n} = n(R-\delta_1)
	= n \omega ~.
\ee
\noindent 
Using eq. (\ref{14}) we then get the correct spectrum of the Calogero $A_{N-1}$
model.

Let us now discuss as to how to obtain the eigenfunctions of CSM using the
formalism of SUSY QM.
We have seen that $A_1$ and $A_1^\dagger$ relate the non-zero eigen states
of the partner Hamiltonians $H$ and $\tilde{H}$. Once a particular state of
$H (\tilde{H})$ with non-zero eigen value is known, the use of eq.
(\ref{21}) enables us to find the corresponding state of 
$\tilde{H}
(H)$. In particular, using eq. (\ref{1.8}) and the fact that in this case 
all the $l_i$ 
are identical, it follows that all
the eigen-functions can be obtained from the ground state wave
functions $\psi_0$ as, $\psi_n=(A_1^\dagger)^n \psi_0$. Note that this
is justified from the operator algebra also, since $A_1$ and $A_1^\dagger$
can be identified as the annhilation and the creation operator respectively.
In particular, one can show
using eqs. (\ref{16}), (\ref{19}), (\ref{23}) and (\ref{24}) that
$[H,A_1]=-A_1$ and $[H,A_1^\dagger]=A_1$. 

This procedure for obtaining the
eigen-functions is similar to that of Isikov {\it et al.} [9].
To see this, define a set of operators,
\be\label{27}
A_n =\sum_{i=1}^N a_i^n, \ \ \ \ n \leq N, \
\ee
\noindent which are symmetric in the particle indices. These operators
satisfy relations which are analogous to those given by eqs. (\ref{16}) and 
(\ref{19}) for any $n$
(see the next paragraph).
It is easily checked that $[H,A_n]=-n A_n$ and
$[H,A_n^\dagger]=n A_n^\dagger$. Following [9], 
the $k$-th eigen-state is given by,
\be\label{28}
\psi_{\{n_i\}} = \prod_{i=1}^N \left ( {A}^\dagger_i \right )^{n_i} \psi_0,
\ \ \ \ a_i \psi_0 =0, \ \ \ \ k=\sum_{i=1}^N n_i ~.
\ee
\noindent Note that $\psi_{\{n_i\}}$ incorporates all the degenerate states
corresponding
to a particular value of $k$ and all the corresponding states of $\tilde{H}$
can be obtained by applying the same $A_1$ on $\psi_{\{n_i\}}$.

Let us now ask the question whether or not $A_1$ is the only operator which
relates the states with nonzero eigen values of the partner Hamiltonians. The
answer obviously is negative and in fact, any operator which is symmetric
in the particle
indices can be used to relate the non-zero eigenstates of the partner
Hamiltonians. However, none of these operators are useful
in deriving the full spectrum of the $A_{N-1}$ CSM model.
For example, if $\psi$ is an
eigen-function of $H$ with
non-zero energy eigen-value $E$, then,
\be\label{29}
\tilde{H} ( A_n \psi ) = \left [ E + \delta_n \right ] ( A_n \psi )~~,
 \ \ \ \ \delta_n = [(N-n)  \pm l N (N-1)] \omega~.
\ee
\noindent Note that the above equation is valid only if $\psi$ is at least
the $n$-th excited state, since $A_n(A_n^\dagger)$ anhilates(creates)
$n$ states.
Similarly, one can show that any state $\tilde{\psi}$ of $\tilde{H}$,
which represents at least the $(n-1)$-th excited state with energy
eigen value $\tilde{E}$, is related to a state
${A_n^\dagger} \tilde{\psi}$ of $H$ with the eigen value 
$\left ( \tilde{E} - \delta_n \right )$. This again proves one to one
correspondence between the $n$-th excited state of $H$ and the $(n-1)$-th
excited state of $\tilde{H}$. As a result, the use of SI gives only
the spectrum beginning with the $n$'th excited state of $H$ and not the 
full spectrum. 

It is worth pointing out that for the $B_N$ type models, however, 
the symmetry arguments force us to replace $A_1$ by $A_2$ in order
to derive the full spectrum using SUSY QM.
This is discussed below in detail.

\subsection{Rational $BC_N$ Calogero Model}

The Hamiltonian for $BC_{N}$ Calogero model is given by
\bea\label{2.1}
H_{BC_{N}} & = & {\frac{1}{2}}[{\sum_{i} {p_{i}^2}}  + 
 l(l \mp 1) \sum_{i,j}^{'} \left [(x_{i} - x_{j}) ^{-2} +
{(x_{i} +x_{j})} ^{-2} \right ]\nonumber \\
 & + & (l_{1}-1)l_{1} {\sum_{i}x_{i}}^{-2}
 + \frac{(l_2)(l_2-1)}{2} \sum_i { x_i}^{-2}+
{\frac{\omega}{2}}  {\sum_{i} x_i^2}] ~.
\eea
\noindent The sign $\mp$ in front of the second term implies
that $H$ is restricted
to act on the space of anti-symmetric (symmetric) wave-functions only.
This model reduces to CSM of $B_N$, $C_N$ and $D_N$
type in the limit $l_2=0$, $l_1=0$ and $l_2=l_1=0$, respectively.
Without loss of generality, in this section, we therefore only study 
the $B_N$ type model,
i.e. $l_2=0$. The other cases are easily obtained from here.

It is interesting to observe that this system also shares  
the property of SI
as found in [7] in the case of the $A_{N-1}$ model. 
Superpotential corresponding to this model is 
\be
W_{i} =\frac{\partial G(x_{1}...x_{N})}{ \partial x_{i}}=
\frac{\partial (ln \psi_0)}{\partial x_i}~,
\ee
\noindent where $\psi_0$ is the ground state wave function of $H_{B_{N}}$
and $G$ is given by 
\be
G = +l_{1} \sum_{i} \ln (x_{i}) +l\sum_{i > j}\ln (x_{i} - 
x_{j}){(x_{i} +x_{j})} -\frac{\omega}{2} \sum_i x_i^2~. 
\ee
\noindent Thus, the superpotential takes the form 
\be
W_i =+l \sum_j^\prime \left [ (x_i - x_j)^{-1} +(x_i +x_j)^{-1} \right ]+
l_{1} x_i^{-1} - \omega x_i~.
\ee
\noindent Following [7], define 
${\cal A}_{i} ({\cal A}_{i}^\dagger)=\pm 
\partial_{i}+W_i $,from which Hamiltonian (\ref{2.1}) with $l_2=0$
can be expressed as 
\be
H^{B_N}=\sum_i {{\cal A}_{i}}^\dagger {\cal A}_i \ \
-\left [ \frac{N}{2} -l N (N-1) -l_{1} N \right] \omega~, \ \
\ee
\noindent
Shape invariance follows due to the
identity,
\be
\sum_{i}{\cal{A}}_{i}{\cal{A}}_{i}^\dagger(l,l_{1}) =
\sum_i {\cal{A}}_{i}^\dagger
{\cal{A}}_{i}(l+1,l_{1} +1)~.
\ee
\noindent
Shape invariance as observed in [7] for $A_{N-1}$ CSM, is
present not only
in the rational $B_N, D_N, BC_N$ models, but also in their trigonometric
counterparts. For trigonometric $BC_N$ models this is shown  
in the Appendix .

As in the $A_{N-1}$ case, the SI condition does not help us in obtaining
the spectrum of the rational $BC_N$ models
unless we employ the
exchange operator formalism. Further, 
the Hamiltonian in eq. (\ref{2.1}) can also be cast in a diagonal form using
exchange
operator method. This however requires including
a reflection operator $(t_{i})$ where $t_{i}$ commutes with $x_{j}$ and 
anti-commutes
with $x_{i}$. The Dunkl derivative operator ( analogous to the $A_{N-1}$ case) 
is given by
\be\label{2.2}
{\cal{D}}_{i}
= -i \partial_{i} + il \sum_{j}^\prime \left [ {(x_i -x_j)}^{-1} M_{ij} +
{(x_i +x_j)}^{-1} \tilde {M_{ij}} \right ] +i l_1 x_{i} ^{-1},\ \ \
\tilde{M_{ij}} = t_{i}t_{j}M_{ij}~.
\ee
\noindent
The reflection operator $t_i$ satisfies the following relations 
\bea
 t_i^2=1~, \ \ t_i \psi(x_1, \dots, x_i, \dots, x_N)=
\psi(x_1, \dots, -x_i, \dots, x_N)~,\nonumber \\
M_{ij} t_i = t_j M_{ij}~,\ \ 
\tilde{M}_{ij}^\dagger = \tilde{M}_{ij}~,\ \
t_i {\cal{D}}_i = - {\cal{D}}_{i} t_i~,\ \
t_i {\cal{D}}_j = {\cal{D}}_j t_i~,\ \ j \neq i,\nonumber \\
\tilde {M_{ij}} {\cal{D}}_i = -{\cal{D}}_{i} \tilde{M_{ij}} \, .
\eea
\noindent It follows from eq. (\ref{2.2}) that $[{\cal{D}}_i,
{\cal{D}}_j] = 0$ and 
\be
[x_i,{\cal{D}}_j]=i\delta_{ij} \left ( 1 + l \sum_{k}^\prime (M_{ik}+
\tilde{M_{ik}})+2l_1t_i \right )
-i(1-\delta_{ij})l(M_{ij} -\tilde{M_{ij}}) ~.
\ee
\noindent
Defining, ${\hat{a}}_i$  and ${\hat{a}}_i^\dagger$ 
with the same defintion as in the previous case (see eq. (\ref{11}))
and using the above equations,
one finds
\be
[{\hat{a}}_i, {\hat{a}}_j^\dagger] = 2 \omega \delta_{ij} \left (1+
l\sum_{k}^{\prime} ( M_{ik}+
\tilde{M_{ik}}) +2l_{1}t_{i} \right )
- 2(1- \delta_{ij}) l \omega (M_{ij} - \tilde{M_{ij}})~.      
\ee

\noindent As before, the SUSY partner Hamiltonians ${\cal{H}}$
and $\tilde {{\cal{H}}}$ for
the $B_{N}$ case 
are defined as 
\be
{\cal{H}} = \frac{1}{2}\sum_{i}{\hat{a}}_i^\dagger {\hat{a}}_i \, ~, \ \ 
\tilde{{\cal{H}}} = \frac{1}{2}\sum_{i}{\hat{a}}_i {\hat{a}}_i^\dagger \, ~.
\ee
\noindent 
It can be seen that,
\be
H_{B_{N}} ={\cal{H}} + E_{0}^{B_{N}}~, \ \ \
E_{0}^{B_{N}} = [\frac{N}{2} \mp \frac{l}{2} N (N-1)+l_1 N] \omega~.
\ee
\noindent
The operator which brings in a correspondence between the 
eigenstates $\psi$
and $\tilde{\psi}$ are respectively
\be
\hat{A}_{2} = \sum_{i} {\hat{a}}_{i}^2 \, ~, \ \
\hat{A}_{2}^\dagger = \sum_i (\hat{a}_i^\dagger)^2 \, ~.
\ee
\noindent One can show that if ${{\psi}} (\tilde {{{\psi}}})$ is the
eigenfunction of 
${\cal{H}} (\tilde{{\cal {H}}})$ with eigenvalue ${\cal{E}}
(\tilde {{\cal{E}}})$ then 
\be
{\cal{H}}(\hat{A_{2}}^\dagger {\psi}) =
(\tilde{{\cal{E}}}-\hat{\delta}_{2})(\hat{A_{2}}^\dagger 
\tilde {\psi})~,\ \
\tilde {{\cal{H}}}(\hat{A}_{2}\psi) =
({\cal{E}}+\hat{\delta}_{2})(\hat{A}_{2}\psi)~~.
\ee
\noindent where,
\be
\hat{\delta}_{2} = [N-2 \pm 2lN(N-1) +2l_{1}N ] \omega~.
\ee
\noindent Now the question is why we
should take $\hat{A}_2$ instead of $\hat{A}_1$ (note that 
$\hat{A}_n=\sum_i \hat{a}_i^n$)?
The point is, unlike the $A_{N-1}$ case, the $BC_N$ Hamiltonian has the 
reflection 
symmetry, $x_{i}{\rightarrow}-x_{i}$. Such a symmetry on the wave-functions 
is ensured only if one uses
$\hat{A}_2$ and not $\hat{A}_1$. 

\noindent Following the treatment in the $A_{N-1}$ case, it is easy to show
that $H$ and $\tilde {H}$ of the $B_N$ model also satisfy the SI condition i.e.
\be
\tilde {H} (\{x_i\},l,l_1) = H (\{x_i\},l,l_1) + R_2 (l,l_1)
\ee
where
\be
R_{2}(l,l_1) =[N  \pm 2lN(N-1) +2l_{1}N] \omega~~.
\ee 
Since in this case also all the $l_i$ are identical, hence it is easy to
see that the spectrum is given by
\be
E_{n}=n(R_{2}-\hat{\delta}_{2})
     =2n\omega~.
\ee     
\noindent  Note that now the spectrum is given by $2n\omega$, instead
of $n\omega$ as
in the case of $\hat{A}_{N-1}$. This spectrum was also obtained earlier in 
[10], but by different method.

Thus we have shown that for the N-body Calogero models,
the spectrum can also be obtained by using the ideas of SQM, SI and exchange 
operator formalism.

\section{Summary \& Discussions }

In this paper, we have generalized the ideas of SUSY QM with one degree
of freedom to the rational-CSM,
which is a many-body problem.  In particular, we have shown that the exchange
operator formalism is suitable for relating the non-zero eigen states
of the partner Hamiltonians of CSM. The shape invariance in this formalism
becomes
trivial compared to the case discussed in [7]. In fact, the
potentials
of the partner Hamiltonians differ by a constant and
this is reminiscent of the usual harmonic oscillator case. 
As a result, the operator method employed in [9] for solving
the rational-CSM algebraically and the SUSY method described here are 
not very different from each other.

One of the nontrivial check of the applicability of the SUSY QM and the SI
ideas to the many-body problems lies in solving the trigonometric CSM,
since unlike the oscillator case, in this case  the energy spectrum is not
linear in the radial quantum number.  Unfortunately, the generalized
momentum operator $D_i$ for all types of models, rational as well as
trigonometric CSM associated with the root structure of $A_n$, $B_n$,
$D_n$ and $BC_n$, are hermitian by construction. So, we can not talk of
partner Hamiltonians in terms of $D_i$ alone. We can define the usual
creation and the anhilation operators, $a_i^\dagger$ and $a_i$, in case we
are dealing with the rational-CSM and construct partner Hamiltonians.
Unfortunately, this can not be done for the trigonometric models. On the
other hand, as described in [7], we can indeed introduce partner
Hamiltonians for $A_n$ type of trignometric models provided the exchange
operator formalism has not been used. The SI is present in this formalism
also but the task of relating the eigenspectrum of the partner
Hamiltonians is unknown as yet.  We have shown in this paper that the SI
is also present in the most general $BC_N$ type of trignometric models.
However, the problem again lies in our inability to relate the the
spectrum of the partner Hamiltonians.

\begin{appendix}

\section{SI in Trigonometric $BC_N$ CSM model}

In this Appendix, we present the SI conditions for the trigonometric
$BC_N$ models.
The trigonometric $BC_N$ Hamiltonian is given by,
\bea 
H_{BC_N} &=& 
- \sum_i \partial_i^2 + l (l - 1) \sum_{i,j}^{\prime} \left [
\frac{1}{\sin^2(x_i - x_j)} + \frac{1}{\sin^2(x_i + x_j)} \right ]\nonumber \\
&& + \sum_i \frac{l_1 (l_1
- 1)} {\sin^2 x_i}
+ l_2 (l_2 - 1) \sum_i \frac{1}{\sin^2 2x_i}~.
\eea
\noindent This model reduces to $B_N$, $C_N$ and $D_N$ for
(a) $l_2=0$, (b) $l_1=0$,
(c) $l_1=l_2=0$, respectively. We define a superpotential
of the form,
\be
W_i = l \sum_j \left [ \cot(x_i - x_j) + \cot(x_i + x_j) \right] + l_1 \cot
x_i + l_2 \cot 2x_i~.
\ee
\noindent 
Using this expression of $W_i$
in the definition of the creation and the anhilation operators as defined
in (\ref{1.1}), one can construct partner Hamiltonians which are equivalent 
to $H_{BC_N}$ up to an overall constant. The SI condition for these partner 
Hamiltonians is
\be 
H_2^{BC_N} (\{x_i\}, l , l_1 , l_2) = 
H_1^{BC_N} (\{x_i\}, l^\prime , l_1^\prime , l_2^\prime) +
R^{BC_N}~,
\ee
\noindent where
\bea
R^{BC_N} & = & 
2 ( l^\prime l_1^\prime - l l_1) N (N - 1) + \frac{4}{3} N (N - 1) (N - 2)
(l^{\prime 2} - l^2) 
+ 4 N (l_2^\prime l_1^\prime - l_2 l_1)\nonumber \\
&& + 4 N (N - 1)
(l_2^\prime l^\prime - l_2 l) + 4 N (l_2^{\prime 2} - l_2)
+ N (l_1^{\prime 2} - l_1) + 2 N (N - 1) (l^{\prime 2} - l^2)
\eea
\noindent and
\be 
l^\prime=l-1~, \ \ l_1^\prime = l_1 - 1 ~,\ \ l_2^\prime=l_2 - 1~.
\ee
\noindent The SI condition for $B_N$, $C_N$ and $D_N$ can be obtained
from the above equations by taking appropriate limits. In particular,
by putting $l_2 = 0$, $l_1 = 0$ or $l_2 = l_1 = 0$, we obtain the corresponding
results for the $B_N$, $C_N$ and $D_N$ type models respectively.\\

\end{appendix}

\end{document}